\documentclass[aps,pra,twocolumn,showpacs,superscriptaddress]{revtex4}
\usepackage{graphicx}
\usepackage{dcolumn}
\usepackage{bm}
\def\om{\omega}

\def\bc{\begin{center}}
\def\ec{\end{center}}
\def\be{\begin{eqnarray}}
\def\ee{\end{eqnarray}}
\def\bea{\begin{eqnarray*}}
\def\eea{\end{eqnarray*}}

\newcommand{\omits}[1]{}
\newcommand\btd{\raise 2pt
\hbox{$\hat\bigtriangledown$}\hskip 1.5pt}
\newcommand\bt{\raise 2pt
\hbox{$\bigtriangledown$}\hskip 1.5pt}

\begin{document}
\title
{Quantum correlated light pulses from sequential superradiance of a
condensate}
\author{M.E. Ta\c{s}g{\i}n}
\affiliation{Department of Physics, Bilkent University, 06800 Bilkent,
  Ankara, Turkey}
\author{M.\"{O}. Oktel}
\affiliation{Department of Physics, Bilkent University, 06800 Bilkent,
  Ankara, Turkey}
\author{L. You}
\affiliation{School of Physics, Georgia Institute of Technology,
Atlanta, Georgia 30332, USA}

\author{\"O. E. M\"{u}stecapl{\i}o\u{g}lu}
\affiliation{Department of Physics, Ko\c{c} University, 34450
Sar\i yer, Istanbul, Turkey}
\date{\today}
\begin{abstract}
We discover an inherent mechanism for entanglement swap associated
with sequential superradiance from an atomic Bose-Einstein
condensate. Based on careful examinations with both analytical
and numerical approaches, we
conclude that as a result of the swap mechanism,
Einstein-Podolsky-Rosen (EPR)-type quantum
correlations can be detected among the scattered light pulses.
\end{abstract}









\maketitle
\section{Introduction}
Superradiance (SR) commonly refers to cooperative emission from an
ensemble of excited atoms with initial coherence or from an
ensemble of radiators with an initial macroscopic dipole moment.
As coherence enhanced radiation, SR was introduced by Dicke
\cite{dicke} in 1954, and first observed experimentally in 1973
\cite{skribanowitz}. It occurs in many systems \cite{gross}, from
thermal gases of excited atoms \cite{vrehen} and molecules
\cite{skribanowitz}, quantum dots and quantum wires
\cite{chen1,chen2,mitra}, to atomic Bose-Einstein condensates
(BEC) \cite{ketterle}, Rydberg gases \cite{yelin}, and molecular
nanomagnets \cite{yukalov}. Recently, serious efforts have been
directed toward the study of quantum entanglement between
condensed atoms and SR light pulses \cite{meystre2000} and
entanglement between atoms through SR \cite{chen2}. Several
promising applications, including prospect for quantum
teleportation in entangled quantum dots via SR, are proposed
\cite{chen2}.

In the pioneering experiment of SR from an elongated condensate, a
continuous wave (cw) pump laser intersects along the short
transverse direction \cite{ketterle}. The scattered radiation is
dominated by axial or the so-called end-fire modes \cite{dicke}.
The atoms experience recoils as a result of the momentum
conservation, exhibiting a fan-like pattern, which reflects the
condensate side--mode distribution. More recently, the
Kapitza-Dirac regime of SR was observed \cite{schneble} in a
pulsed pump scheme, with momentum side--modes displaying the
characteristic X-shaped patterns. In this regime, it is predicted
that SR pulses must contain quantum entangled counter-propagating
photons from the end-fire modes \cite{meystre2003}. It was
proposed that quantum entanglement arises from correlations of
backward and forward scattered atoms and from the interplay
between optical and atomic fields \cite{meystre2003}. In this
study we show that even for a cw pumped condensate with scattered
atoms forming a forward fan-like pattern, quantum entanglement of
the end-fire modes still exists due to an entanglement swap
mechanism which we clearly identify during sequential SR process.
In quantum information language, entanglement swap is a technique
to entangle particles that never before interacted
\cite{zukowski,bennett,bose,pan}.

Sequential SR involves successive scattering of the pump laser
from the initial momentum distribution of a condensate
\cite{ketterle}. Previous studies on SR from an atomic gas have
observed multiple pulses or ringing effects, especially among
dense atomic samples. Ringing is often explained in terms of the
pulse propagation effect \cite{bonifacio75}, where the finite size
and shape of the medium plays significant roles
\cite{arecchi,rehler}. Adopting semi-classical theories, detailed
modeling of SR from atomic condensates have been very successful,
essentially capable of explaining both spatial and temporal
evolutions of atomic and optical fields
\cite{zobay,robb,benedek,vasilev}. The semi-classical treatments,
however, can account neither for the influence on sequential
scattering associated with ring from side--mode patterns nor for
quantum correlations between end-fire modes.

In this paper, we investigate Einstein-Podolsky-Rosen (EPR)-type
\cite{epr,pu} quantum correlations between end-fire modes. Such
correlations can be detected with well known methods developed
for continuous variable entanglement in down-converted
two-photon systems \cite{howell,dangelo},
employing equivalent momentum and position quadrature variables
as observable.

The paper is organized as follows. In sec.
\ref{sec:Effective-Hamiltonian}, we introduce the
relevant concepts and describe the model system
we consider for investigating sequential SR.
We identify the various approximations and
derive the full second quantized effective Hamiltonian.
In sec.
\ref{sec:entanglement-criteria}, we review the
criteria for continuous variable entanglement,
with which we confirm the existence of quantum
correlation between SR photons from the end-fire modes.
In sec.
\ref{sec:swapmechanism}, we analytically
solve the effective Hamiltonian
under parametric and steady state approximations.
We clearly identify the swap mechanism, and
intuitively explain the steps involved for the
model Hamiltonian to generate EPR pairs out of non-interacting
photons. This represents the key result for this article.
In sec.
\ref{sec:calculation}, we describe the method of our numerical
calculations under a proper decorrelation approximation.
The results are
presented in sec. \ref{sec:results}, where we first examine the temporal
dynamics of the entanglement in connection with the accompanying
field and atomic populations. This helps to illustrate the swap
of atom-photon entanglement to the photon-photon entanglement.
We then study carefully this swap effect, introduce the effect of
decoherence, and consider the effect of SR initialization
from a two-mode squeezed vacuum and the dependence on
the increase/decrease of number of atoms.
Sec. \ref{sec:conclusions} contains our conclusion.
\section{Sequential Superradiance and the Effective Hamiltonian}
\label{sec:Effective-Hamiltonian}
In this section, we briefly describe the unique properties of SR,
{\it e.g.}, the directional and sequential nature of the emitted
pulses. We will introduce the concept of sequential SR, in terms
of what occurs in an elongated cigar-shaped atomic condensate. We
derive the second quantized effective Hamiltonian, where the
optical fields are treated quantum mechanically, in order to take
into account the interaction of all side--modes with a common
photonic fields.

\subsection{Sequential SR}

We consider an elongated condensate, of length $L$ and width $W$,
that is  axially symmetric with respect to the long direction of the $z$-axis.
It is optically excited with a strong pump laser of frequency
$\omega_0$, detuned from the atomic resonance frequency $\om_A$ by
$\Delta=\om_A-\om_0$. The laser beam is directed along the $y$-axis,
perpendicular to the long axis of the condensate, and linearly
polarized in the $x$-axis.

When the pump laser is sufficiently strong, the occupation of
atoms in the excited state
becomes macroscopic, beyond the threshold for collective
emission. The excited atoms, interacting through the
common electromagnetic field, start to make collective spontaneous
emission \cite{dicke}. During this dynamic process, the uncorrelated state of
our model system first evolves linearly
into a correlated state in the early stages
with the initiation of a superradiant pulse from vacuum noise.
Subsequently this is followed by a nonlinear regime where
the superradiant pulse is fully developed and its
interaction with atoms becomes completely nonlinear.
The decay of the pulse then occurs in a final linear dynamic stage.
At the peak of SR, the collective radiation time of the system
$T_R=\left(8\pi/n\lambda^2 L\right)T\sim10^{-3}T\sim10^{-10}$ s,
 becomes much smaller than the
normal spontaneous emission time $T\sim 60$ ns for typical
systems, where $n$ is the
density of atoms in the excited state, $\lambda$ is the
resonant transition wavelength.

\begin{figure}
\includegraphics[width=3.35in]{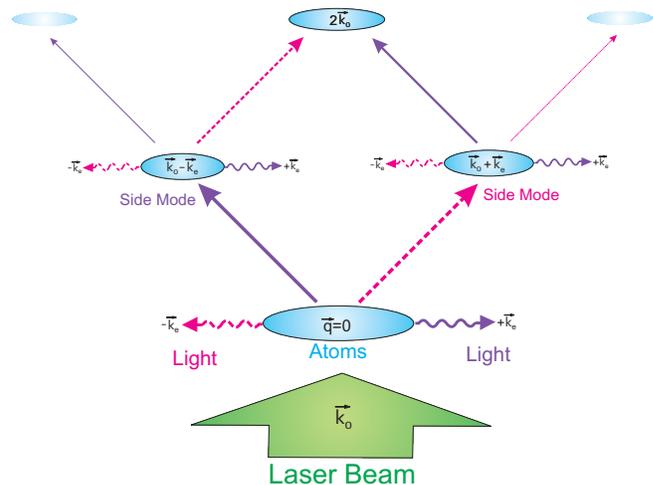}
\caption{(Color online) A fan-like atomic
side--mode pattern up to second order sequential superradiant scattering. }%
\label{fig1}
\end{figure}

For an elongated radiating sample, as the condensate along the
$z$-axis being discussed here, superradiant emission occurs
dominantly along the $\pm\hat{z}$ directions, {\it i.e.}, emitted
photons leaving the cigar shaped sample mainly from both ends. The
corresponding spatial modes are called end-fire modes. They are
perpendicular to the propagation direction of the pump laser beam.
Due to momentum conservation for individual scattering events, the
emission of end-fire photon is accompanied by collective recoils
of condensate atoms. The momentum of recoiled atoms is
significantly larger in magnitude than the residue momentum spread
of the trapped condensate, thus gives rise to distinctly different
islands of clouds upon on free expansion. These are the so-called
condensate side--modes. When the side--modes are occupied
significantly, they serve as new sources for higher order SR, or
sequential SR. They, too, emit end-fire mode photons and
contribute to the next order side--modes. The resulting pattern
for atomic distribution after expansion as depicted in Fig.
\ref{fig1}, corresponds to what was observed for a certain choice
of pump power and duration in the first BEC SR experiment
\cite{ketterle}. The directions of the emitted end--fire mode
photons and the corresponding recoiled side--mode condensate
bosons are indicated with the same line type.
%
%
\subsection{Effective Hamiltonian}
The effective second quantized Hamiltonian, governing the dynamics of
sequential SR system, is derived as follows. Due to the large
energy scale difference between the center of mass (CM) dynamics
for the atoms ($\sim$MHz) and the internal electronic degrees of
freedom ($\sim$PHz), we can treat separately their respective motions.
As in Ref. \cite{meystrePRA}, the Hamiltonian of an atomic condensate
with two level atoms interacting with a near-resonant laser pump,
takes the following form
\begin{eqnarray}
\hat{H} &=& \int d^3\mathbf{r} \hat{\psi}_g^{\dagger}(\mathbf{r})
\left( -\frac{\hbar^2}{2m} \nabla^2 + V_{\rm tg}(\mathbf{r}) \right)
\hat{\psi}_g(\mathbf{r}) \nonumber \\
 & &+ \int d^3\mathbf{r}
\hat{\psi}_e^{\dagger}(\mathbf{r}) \left( -\frac{\hbar^2}{2m}
\nabla^2 + V_{\rm te}(\mathbf{r}) +\hbar\omega_a\right)
\hat{\psi}_e(\mathbf{r}) \nonumber \\
& &+ \int d^3\mathbf{k} \hbar\omega_k \hat{a}_\mathbf{k}^{\dagger}
\hat{a}_\mathbf{k} \nonumber \\
&&+ \int d^3\mathbf{r} d^3\mathbf{k} \left[ \hbar g^*(\mathbf{k})
e^{-i\mathbf{k}\cdot\mathbf{r}} \hat{\psi}_g^{\dagger}(\mathbf{r})
\hat{a}_\mathbf{k}^{\dagger}
\hat{\psi}_e(\mathbf{r}) 
\: + {\rm H.c.} \right], \hskip 12pt
 \label{eq:Hamilt1}
\end{eqnarray}
under the dipole and rotating wave approximations. We have further
neglected the static atom-atom interactions. The first two terms are
the atomic Hamiltonians for the CM motion in their respective
trapping potentials $(V_{\rm tg}(\mathbf{r})$,\,$V_{\rm
te}(\mathbf{r}))$ of the internal states. The atomic fields,
described by annihilation (creation) operator
$\hat{\psi}_{g,e}(\mathbf{r})$
 ($\hat{\psi}^\dagger_{g,e}(\mathbf{r})$), obey the usual bosonic
algebra. $E_g=0$ and $E_e=\hbar\Delta=\hbar(\omega_A-\omega_0)$ are
the electronic excitation energies of the atom in the rotating
frame defined by the pump laser field.
The third term comes from the free electromagnetic field,
while its interaction with the atoms is described with the last term
($\hat{H}_{af}$), which includes both the
laser photons and the scattered photons. The operator $\hat{a}_{\mathbf{k}}$
($\hat{a}^\dagger_{\mathbf{k}}$) annihilates (creates) a photon with
wavevector $\mathbf{k}$, polarization $\hat{\epsilon}_\mathbf{k}$,
and frequency $\omega_\mathbf{k}=ck-\omega_0$ (again in the rotating
frame with frequency $\omega_0$).
$g(\mathbf{k})=\left[c|\mathbf{k}|d^2/2\hbar\epsilon_0(2\pi)^3\right]^{1/2}
|\hat\mathbf{k}\times\hat\mathbf{x}|$ is the dipole coupling
coefficient, with $\vec{d}=\langle e|\vec{r}|g \rangle$ the
matrix element for the atomic dipole transition.

In typical SR experiments \cite{ketterle}, the detuning $\Delta\sim
10^9\,$Hz is much larger than both the CM motion energy scale
$\sim10^6\,$Hz or the Rabi frequency $\Omega_0\sim10^8\,$Hz, the
excited state operator can be eliminated adiabatically via replacing
$\hat{\psi}_e(\mathbf{r})\approx-(1/\Delta) \left( \int
d\mathbf{k}g(\mathbf{k}) e^{i\mathbf{k}\cdot\mathbf{r}}
\hat{a}_\mathbf{k} \right) \hat{\psi}_g(\mathbf{r})$ in the
equations of motion, yielding an effective Hamiltonian
\begin{eqnarray}
\hat{H} &=&  \int d^3\mathbf{r} \hat{\psi}_g^{\dagger}(\mathbf{r})
\left( -\frac{\hbar^2}{2m} \nabla^2 + V_{\rm tg}(\mathbf{r}) \right)
\hat{\psi}_g(\mathbf{r}) \nonumber\\
&+& \int d^3\mathbf{k}
\hbar\omega_k \hat{a}_\mathbf{k}^{\dagger} \hat{a}_\mathbf{k} \nonumber \\
&-&\frac{\hbar}{\Delta} \int d^3\mathbf{r} d^3\mathbf{k}
 d^3\mathbf{k}'\tilde{g}(\mathbf{k},\mathbf{k}',\mathbf{r})
\hat{\psi}^{\dagger}_g(\mathbf{r}) \hat{a}^{\dagger}_{\mathbf{k}}
\hat{a}_{\mathbf{k}'} \hat{\psi}_g(\mathbf{r}), \label{eq:Hamilt2}
\end{eqnarray}
with $\tilde{g}(\mathbf{k},\mathbf{k}',\mathbf{r})=g^*(\mathbf{k})
g(\mathbf{k}') \exp{({-i(\mathbf{k}-\mathbf{k}')\cdot
\mathbf{r}})}$, proportional to the effective coupling between
the absorbed and subsequently emitted photons.

The atomic field operators can be expanded in terms of the
quasi-particle excitations of BEC
$\hat{\psi}_g(\mathbf{r})=\sum_\mathbf{q}\langle\mathbf{q}|\mathbf{r}\rangle
\hat{c}_\mathbf{q}$, as described in Ref. \cite{meystre}, with
$\hat{c}_\mathbf{q}$ ($\hat{c}^\dagger_\mathbf{q}$) annihilating
(creating) a scattered boson in the momentum side--mode
$\mathbf{q}$ in the form
$\langle\mathbf{r}|\mathbf{q}\rangle=\phi_0(\mathbf{r})
e^{i\mathbf{q}\cdot\mathbf{r}}$. The initial condensate mode is
described by the spatial wave function $\phi_0(\mathbf{r})$. The
quasi modes for excitations approximately form an orthonormal
basis because $\langle
\mathbf{q}|\mathbf{q}'\rangle=\delta_{\mathbf{q},\mathbf{q}'}$. In
the second quantized form within the side--mode representation Eq.
(\ref{eq:Hamilt2}) becomes
\begin{eqnarray}
\hat{H}&&=\sum_{\mathbf{q}} \hbar\omega_q
\hat{c}_{\mathbf{q}}^\dagger \hat{c}_{\mathbf{q}} + \int
d^3\mathbf{k} \hbar\omega_k \hat{a}_\mathbf{k}^{\dagger}
\hat{a}_\mathbf{k} \nonumber \\%
 &&-\frac{\hbar}{\Delta} \sum_{\mathbf{q},\mathbf{q}'} \int d^3\mathbf{k} d^3\mathbf{k}'
g^*(\mathbf{k}) g(\mathbf{k}')
\rho_{\mathbf{q},\mathbf{q}'}(\mathbf{k},\mathbf{k}')
\times\nonumber\\ &&\times \hat{c}_{\mathbf{q}}^{\dagger}
\hat{a}_{\mathbf{k}}^{\dagger} \hat{a}_{\mathbf{k}'}
\hat{c}_{\mathbf{q}'}, \label{eq:Hamilt3}
\end{eqnarray}
where $\rho_{\mathbf{q},\mathbf{q}'}(\mathbf{k},\mathbf{k}')=\int
d\mathbf{r} |\phi_0(\mathbf{r})|^2
e^{i[(\mathbf{k}+\mathbf{q})-(\mathbf{k}'+\mathbf{q}')]\cdot\mathbf{r}}$
is the structure form factor of the condensate density, which is
responsible for the highly directional emission of the end-fire
mode photons. $\omega_\mathbf{q}=\hbar|\mathbf{q}|^2/2m$ is the
side--mode energy at the recoil momentum of $\mathbf{q}$. The
first two terms in Eq. (\ref{eq:Hamilt3}) are diagonal in their
respective Fock spaces and can be omitted by performing further
rotating frame transformations $\hat{c}_\mathbf{q} \rightarrow
\hat{c}_\mathbf{q}e^{-i\omega_\mathbf{q}t}$ and
$\hat{a}_\mathbf{k} \rightarrow
\hat{a}_\mathbf{k}e^{-i\omega_\mathbf{k}t}$. Thus, the effective
Hamiltonian takes the form
\begin{eqnarray}
\hat{H} = -\frac{\hbar}{\Delta} \sum_{\mathbf{q},\mathbf{q}'} \int
d^3\mathbf{k} d^3\mathbf{k}' g^*(\mathbf{k}) g(\mathbf{k}')
\rho_{\mathbf{q},\mathbf{q}'}(\mathbf{k},\mathbf{k}') \times \nonumber \\
\times \hat{c}_{\mathbf{q}}^{\dagger} \hat{a}_{\mathbf{k}}^{\dagger}
\hat{a}_{\mathbf{k}'} \hat{c}_{\mathbf{q}'}
e^{i\left(\omega_\mathbf{k}+\omega_q
-\omega_{\mathbf{k}'}-\omega_{q'}\right)t}.
\label{eq:Hamilt4}
\end{eqnarray}

In a sufficiently elongated condensate, large off-axis Rayleigh
scattering is suppressed with respect to the end-fire modes
\cite{mustecap}. The angular distribution of the scattered light
is sharply peaked at the axial directions ($\mathbf{k}_e=\pm
k_e\hat{z}$), if the Fresnel number is larger than one ${\cal
F}=W^2/L\lambda_0 \gtrsim 1$ at the pump wavelength $\lambda_0$
for a condensate of length $L$ and width $W$ \cite{meystre}. This
makes it possible to consider only the axial end-fire modes. To
investigate sequential SR, we further take into account the first
order side--modes at $\mathbf{q}=\mathbf{k}_0\pm\mathbf{k}_e$ and
the second order side--mode at $\mathbf{q}\approx2\mathbf{k}_0$.
The rest of the side--modes are assumed to remain unpopulated
\cite{zobay}. The Hamiltonian (\ref{eq:Hamilt4}) that originally
contains the contributions from all the side--modes and the
end-fire modes as well as the laser field then reduces to the
following simple model
\begin{eqnarray}
\hat{H} &=& -\hbar\frac{g^2}{\Delta}\left( \hat{c}_+^{\dagger}
\hat{a}_-^{\dagger} \hat{a}_0 \hat{c}_0 + \hat{c}_-^{\dagger}
\hat{a}_+^{\dagger} \hat{a}_0 \hat{c}_0\right.\nonumber\\
&&\hskip 30pt \left.+\hat{c}_2^{\dagger}\hat{a}_-^{\dagger}
\hat{a}_0 \hat{c}_-+\hat{c}_2^{\dagger} \hat{a}_+^{\dagger}
\hat{a}_0
 \hat{c}_+ \right) +{\mathrm H.c.},
\label{eq:hamiltonian}
\end{eqnarray}
with $g\equiv g(\mathbf{k_e})$. We have adopted a shorthand notation
where
 $\hat{a}_\pm \equiv \hat{a}_{\pm\mathbf{k}_e}$,
$\hat{a}_0 \equiv \hat{a}_{\mathbf{k}_0}$, $\hat{c}_\pm \equiv
\hat{c}_{(\mathbf{k}_0\pm\mathbf{k}_e)}$, and $\hat{c}_2 \equiv
\hat{c}_{2\mathbf{k}_0}$. This is the model Hamiltonian involving
the interplay of the four atomic side--modes with three photonic
modes. Before we further discuss and reveal the built-in
entanglement swap mechanism for EPR-type quantum correlations in
this model Hamiltonian, in the next section we shall briefly
review continuous variable entanglement and extend its criteria to
our case.
\section{Criteria for Continuous Variable Entanglement} \label{sec:entanglement-criteria}

The existence of continuous variable entanglement
is determined by a sufficient condition on the
inseparability of continuous variable states as given in Ref.
\cite{duan2000}.
If the density matrix of a quantum system is inseparable within
two well defined modes \cite{duan2000,ping}, or the two modes are
entangled, the total variance of EPR-like operators,
$\hat{u}=|c|\hat{x}_1+\hat{x}_2/c$ and
$\hat{v}=|c|\hat{p}_1-\hat{p}_2/c$, satisfies the inequality
\begin{equation}
\langle \Delta \hat{u}^2\rangle + \langle \Delta \hat{v}^2\rangle <
\left(c^2 +1/c^2\right),
\label{eq:inseperability}
\end{equation}
for a real number $c$, where $\hat{x}_{1,2}=( \hat{a}_{1,2} +
\hat{a}_{1,2}^\dagger)/\sqrt{2}$ and $\hat{p}_{1,2}=( \hat{a}_{1,2}
- \hat{a}_{1,2}^\dagger)/i\sqrt{2}$ are analogous to position and
momentum operators as in the case of a simple harmonic oscillator.
The indices corresponds to mode numbers.

Defining the inseparability parameter
\begin{equation}
\lambda=\langle \Delta \hat{u}^2\rangle + \langle \Delta
\hat{v}^2\rangle - \left(c^2 +1/c^2\right), \label{eq:lambda}
\end{equation}
the presence of continuous variable entanglement
is then characterized by the sufficient condition $\lambda<0$.
For the two modes to be entangled, it suffices to find only one
value of $c$ that leads to $\lambda<0$, and hence $c$
can be taken at
which $\lambda$ is minimum.

The parameter $\lambda$ we adopt here clearly
corresponds to an entanglement witness, but not an
entanglement measure. Because the states for more negative $\lambda$  do not
necessarily correspond to more entangled states.

The total variance of the EPR operators are bounded below by the
Heisenberg uncertainty relation $\langle \Delta \hat{u}^2\rangle +
\langle \Delta \hat{v}^2\rangle \ge \left| c^2 -1/c^2\right|$. Thus,
$\lambda$ has a lower bound
$\lambda_{\text{low}}=\left|c^2-1/c^2\right|-\left(c^2+1/c^2\right)$.

After minimization with respect to $c$, a more explicit expression
of $\lambda$ can be given as
\begin{eqnarray}
\lambda=&&2\left( c^2\langle \hat{a}^\dagger_1 \hat{a}_1 \rangle +
\langle \hat{a}^\dagger_2 \hat{a}_2 \rangle/{c^2}
+\text{sign}(c)\langle \hat{a}_1 \hat{a}_2 +
\hat{a}^\dagger_1 \hat{a}^\dagger_2 \rangle  \right)\nonumber\\
&&-\langle\hat{u}\rangle^2-\langle\hat{v}\rangle^2,
\label{eq:lambda-endfires}
\end{eqnarray}
where $c^2=\left[(\langle \hat{a}^\dagger_- \hat{a}_-
\rangle-|\langle \hat{a}_-\rangle|^2) / (\langle \hat{a}^\dagger_+
\hat{a}_+ \rangle-|\langle \hat{a}_+\rangle|^2)\right]^{1/2}$ with
$\text{sign}(c)=-\text{sign}\left[\text{Re}\left\{\langle\hat{a}_+\hat{a}_-\rangle\right\}-
\alpha_+\alpha_-+\beta_+\beta_- \right]$,
$\alpha_\pm=\text{Re}\left\{ \langle \hat{a}_\pm \rangle\right\}$
and $\beta_\pm=\text{Im}\left\{ \langle \hat{a}_\pm
\rangle\right\}$.

In the system we study here, the bosonic mode operators will be
either end-fire mode pairs $a_{1,2}=a_{\pm}$ or end-fire modes and
first side--modes, $a_{1}=a_{\pm},a_2=c_{\mp}$. Unlike other model
investigations \cite{ping} of EPR-like correlations based upon
$\lambda$, we need to keep track of the $\langle\hat{u}\rangle^2$
and $\langle\hat{v}\rangle^2$ terms, because
$\langle\hat{x}_{1,2}\rangle$ and $\langle\hat{p}_{1,2}\rangle$ do
not necessarily vanish for our model during time evolution.
Furthermore, since the time evolution of the two end-fire modes
are symmetric in our case, we find $c^2=1$ and
$\lambda_{\text{low}}=-2$.

In the remainder of this paper, we examine the time evolution of
the continuous variable entanglement witness $\lambda(t)$ both for
the opposite end-fire modes and for the end-fire modes with
side--modes. This study is expected to provide insight into the
temporal development and the swap of quantum correlations between
different sub-systems/modes. The following section is aimed at
establishing an intuitive understanding of how EPR-like
correlations between opposite end-fire modes are built up.
\section{An entanglement Swap Mechanism} \label{sec:swapmechanism}
In sec. \ref{sec:results} we will exhibit the numerical results
for the time evolution of the entanglement parameter $\lambda(t)$,
governed by the Hamiltonian (\ref{eq:hamiltonian}). We will
observe that, there exists regions in time where $\lambda$ becomes
negative, {\it i.e.}, conclusive evidence for the presence of
entanglement during dynamical evolution. In this section, we hope
to provide an intuitive understanding to support the result
revealed through the numerical approach. We will show that it is
due to the presence of an inherent swap mechanism which leads to
the generation of the EPR photon pair. We shall examine the
dynamical behavior of the system in two different time regimes,
the early times when the first side--modes just start to grow and
the later times when the second order side--mode contributes to
the dynamics.
\subsection{Early Times}

In the initial stage, occupation of the second order side--mode
($|c_2\rangle$) can be neglected. During this initiation period of
the short-time dynamics, the number of atoms in the zero-momentum
state can be assumed undepleted
$\hat{c}_0\approx\sqrt{N}e^{i\phi_1}$ with a constant $N$ standing
for the number of condensed atoms, like in the treatment of
degenerate parametric processes. Since the pump is very strong,
and the number of pump photons are much larger than the number of
condensate atoms $M\gg N$, it can also be treated within the
parametric pump approximation $\hat{a}_0 \approx \sqrt{M}
e^{i\theta_0}$ as undepleted. Thus, the initial behavior of the
system is governed by the Hamiltonian
\begin{eqnarray}
\hat{H}_1 = -\hbar\chi_1
\left[e^{i\theta_1}\left(\hat{a}_+^{\dagger} \hat{c}_-^{\dagger} +
\hat{a}_-^{\dagger} \hat{c}_+^{\dagger}\right) + {\rm H.c.}
\right], \label{eq:H_early}
\end{eqnarray}
with $\chi_1=\sqrt{NM}|g|^2/\Delta$ and $\theta_1=\theta_0+\phi_1$
is the initial phases difference. This form of $\hat{H}_1$ is
exactly the same as that of two uncoupled optical parametric
amplifiers (OPAs). It allows for the growth of the first-order
side--modes \cite{meystre} as well as the entanglement of
side--mode atoms with the end-fire mode photons \cite{paris}. The
solution to $\hat{H}_1$ in the Heisenberg picture is given by the
following time dependencies of operators \cite{mandel}
\begin{eqnarray}
& \hat{a}_\pm(t)=\text{cosh}(\chi_1 t) \hat{a}_\pm + ie^{i\theta}
\text{sinh}(\chi_1 t) \hat{c}_\mp^\dagger , &
\label{eq:sol_early1} \\
& \hat{c}_\pm(t)=\text{cosh}(\chi_1 t) \hat{c}_\pm + ie^{i\theta}
\text{sinh}(\chi_1 t) \hat{a}_\mp^\dagger ,& \label{eq:sol_early2}
\end{eqnarray}
where the operators without time arguments are at the initial time.

The side--modes and the end-fire modes are initially unoccupied
$|a_+,a_-,c_+,c_-\rangle=|0,0,0,0\rangle$, or taken to be in their
respective vacuum states. The time dependencies for the
populations of the side--modes and end-fire modes come out as
$\langle\hat{I}_{\pm}\rangle=\langle\hat{n}_\pm\rangle=\text{sinh}^2(\chi_1
t)$, analogous to the classical results \cite{mandel}. Evaluating
correlations between the two end-fire modes (e), we find
\begin{equation}
\lambda\equiv\lambda_{\text{ee}}=4\text{sinh}^2(\chi_1 t)  ,
\label{eq:lamdaee-early}
\end{equation}
which is always positive $\lambda_{\text{ee}}>0$. On the other
hand, the correlation between the end-fire mode (e) and side--mode
(s), scattered in the opposite directions, takes the following
form
\begin{equation}
\lambda_{\text{se}}=2\left[2\text{sinh}^2(\chi_1
t)-|\text{sin}(\theta_1)|\text{sinh}(2\chi_1 t) \right]  ,
\label{eq:lamdaes-early}
\end{equation}
which starts with $\lambda_{\text{se}}(t)=0$ and evolves down to
$\lambda_{\text{se}}(t_0 \gtrsim 2)\simeq-2$, the lowest possible
value of $\lambda_{\text{low}}=-2$ imposed by the Heisenberg
uncertainty, at $|\sin(\theta_1)|\simeq 1$.

In Figs. \ref{fig3}(a) and \ref{fig3}(b), this is further
supported by more elaborate results from numerical calculations of
$\lambda$ and $\lambda_{\text{se}}$. Same conclusions can be seen
for the analytical results (\ref{eq:lamdaee-early}) over the range
$t=0$-$t=0.3$ms in Fig. \ref{fig3}(a) and (\ref{eq:lamdaes-early})
over the range $t=0$-$0.2$ms in Fig. \ref{fig3}(b).

\subsection{Later Times}

At later times, the first-order side--modes become effectively
populated, giving rise to noticeable second sequence of SR from
the edges of these side--mode condensates. In this case, the
occupancy for the $|c_0\rangle$ mode is not an important issue,
but the $|c_2\rangle$ mode becomes populated due to the second
order SR.

We construct an approximate model by assuming that the
occupation of $|c_2\rangle$ is not changing too much,
or effective treating it as in in the steady state with
$\hat{c}_2\approx\sqrt{N_2}e^{-i\phi_2}$. $N_2$ is the number of
atoms in the $|c_2\rangle$ state. The later stage dynamics
of the system, where the
$\text{2}^{\text{nd}}$ order SR is effective, is then governed by the
model Hamiltonian
\begin{eqnarray}
\hat{H}_2 = -\hbar\chi_2
\left[e^{i\theta_2}\left(\hat{a}_-^{\dagger} \hat{c}_- +
\hat{a}_+^{\dagger} \hat{c}_+\right) + {\rm H.c.} \right],
\label{eq:H_later}
\end{eqnarray}
with $\chi_2=\sqrt{N_2M}|g|^2/\Delta$ and
$\theta_2=\bar{\theta}_0+\phi_2$. As before we again neglect the depletion
of the pump $\hat{a}_0\approx\sqrt{M}e^{i\bar{\theta}_0}$.

This model $\hat{H}_2$ is also exactly solvable. The time dependencies of the
annihilation operators in the Heisenberg picture are
\begin{eqnarray}
\hat{a}_\pm(t)&=&\text{cos}(\chi_2\Delta t) \hat{a}_\pm +
ie^{i\phi_2}
\text{sin}(\chi_2\Delta t) \hat{c}_\pm,\label{eq:sol_later1} \\
\hat{c}_\pm(t)&=&\text{cos}(\chi_2\Delta t) \hat{c}_\pm + i
e^{-i\phi_2} \text{sin}(\chi_2\Delta t) \hat{a}_\pm,
\label{eq:sol_later2}
\end{eqnarray}
where $t>t_0$, the operators without time arguments are at
$t=t_0$, and $\Delta t=t-t_0$. We can approximately connect these
two models together into smooth temporal dynamics if we use the
solutions of $\hat{H}_1$ as the initial state for dynamics due to
$\hat{H}_2$ so that $\hat{a}_\pm(t_0)$ and $\hat{c}_\pm(t_0)$ are
calculated at $t=t_0$ from the equations (\ref{eq:sol_early1}) and
(\ref{eq:sol_early2}), respectively. We define $t_0$ as the time
at which all the $|c_0\rangle$ atoms are scattered into the
side--modes, and thus it is determined from $\text{sinh}^2(\chi_1
t_0)=N/2$.

In this later dynamical stage, the entanglement witness parameter
in between the end-fire modes (e) is evaluated to be
($\lambda\equiv\lambda_{\text{ee}}$)
\begin{eqnarray}
\lambda(t)&=&4\text{sinh}^2(\chi_1 t_0)\nonumber\\
&-&|\text{cos}(\bar\theta)\text{sin}(2\chi_2\Delta t)|
\text{sinh}(2\chi_1 t_0), \label{eq:lambdaee-later}
\end{eqnarray}
where $\bar\theta=\theta_1+\theta_2$. When
$|\cos(\bar\theta)|\simeq 1$, $\lambda$ evolves from $2N$ down to
the minimum possible negative value of $\lambda_{\text{low}}=-2$,
at $\Delta t=\pi/4\chi_2$. An analogous calculation for
entanglement between the end-fire mode (e) and side--mode (s)
gives
\begin{eqnarray}
\lambda_{\text{se}}(t)&=&4\text{sinh}^2(\chi_1 t_0)\nonumber\\
&-&2|\text{sin}(\theta_2)\text{cos}( 2\chi_2\Delta t)|
\text{sinh}(2\chi_1 t_0),\label{eq:lambdaes-later}
\end{eqnarray}
which starts at $\lambda_{\text{se}}(t_0)=-2$ and increases to
values of order $\sim N$, for proper choices of $\theta_2$. Many
of these features revealed from simple analytic models find their
parallels in numerical solutions as displayed in Fig. \ref{fig3}.

The results from the two model Hamiltonians are found to depend on
the initial phase difference between $\theta_1$ and $\theta_2$, but
not the individual phases. Such a phase dependence of the results is
analogous to the cases of parametric down conversion and the
two-mode squeezing \cite{mandel}. The phases introduced in the
second stage reflects the accumulating temporal phase difference of
the operators through the time evolution. In the numerical
calculation it is sufficient to assign initial phases for the pump
laser and the condensate or just their difference.

Without any detailed analysis, simply consider the behaviors of
Eqs. (\ref{eq:lamdaes-early}) and (\ref{eq:lambdaee-later})
instead, one can already appreciate the built-in entanglement swap
mechanism within the superradiant BEC in action. The entanglement
created between the side--mode and end-fire mode Eq.
(\ref{eq:lamdaes-early}) in the initial stage, is swapped to
entanglement between the two end-fire modes Eq.
(\ref{eq:lambdaee-later}), due to the $\text{2}^{\text{nd}}$ order
SR. The model Hamiltonian $\hat{H}_1$ couples the $|a_\pm\rangle
\leftrightarrow |c_\mp\rangle$ modes, but leaves $|a_+\rangle
\leftrightarrow |a_-\rangle$ modes decoupled at the initial times.
The model Hamiltonian $\hat{H}_2$, at later times, couples the
$|a_\pm\rangle \leftrightarrow |c_\pm\rangle$ states. Two
noninteracting modes $|a_+\rangle \leftrightarrow |a_-\rangle$ are
coupled through their common interaction with the same side--mode,
and become entangled due to the swap mechanism.

\section{Numerical Calculation of the Entanglement Parameter} \label{sec:calculation}
We study the dynamics of the entanglement parameter $\lambda(t)$
and the accompanying populations for the fields
($I_0(t)$,$I_\pm(t)$) and the atomic states
($n_0$,$n_\pm(t)$,$n_2$). Their complete temporal evolution is governed by
the Hamiltonian Eq. (\ref{eq:hamiltonian}). Our calculation will be
numerically obtained, aided by a decorrelation approximation that neglects higher order
correlations. The numerical results will be illustrated and discussed
in the next section.

The entanglement parameter $\lambda$, given in Eq.
(\ref{eq:lambda-endfires}), is determined by
the expectation values of
both $\hat{a}_\pm$ operators and their products.
Their equations of motion in operator forms can be derived
from the full Hamiltonian Eq. (\ref{eq:hamiltonian}).
The dynamics of two operator products is found to depend
on four operator products;
and four operator products depend on six
operator products, and so on so forth.
Such a hierarchy of operator equations
is impossible to manage in general.We therefore resort to a
decorrelation approximation that truncates it to a closed form. The
usual treatment of this kind \cite{andreev} for the SR system closes
the chain early by a simple decorrelation of atomic and optical
operators, which is clearly inappropriate when entanglement swap is
to be studied.

We adopt a decorrelation rule that factorizes condensate and the
second order side--mode operators in operator products. Since
quantum correlations between the condensate and other modes are
expected to be weak due to the almost classical,
coherent-state-like nature for the condensate and its diminishing
population when the second order side--mode is significantly
populated at later stages of dynamics. Operators for the pump
photons will also be factorized, again relying on the almost
classical, coherent state nature of the pump field.

Our approach makes it possible to keep quantum correlations
between the end-fire modes and the intermediate side--modes. The
hierarchy of equations is closed under $\langle xyz\rangle\approx
\langle xy\rangle\langle z\rangle$, with
$x,y\in\{1,c_\pm,c_\pm^\dag,a_\pm,a_\pm^\dag\}$ and
$z\in\{c_0,c_0^\dag,c_2,c_2^\dag\}$. The resulting equations,
governing the dynamics of the expectations, are given in the
Appendix. These equations are solved numerically.

For the initial conditions, both the end-fire modes and
side--modes are taken to be their vacuum Fock states while the
laser and the condensate are in coherent states. We consider a
system with typical parameters of a condensate with number of
atoms $N=8\times10^6$ and a pump with $M=2\times10^8$ photons.
Additionally, phenomenological decoherence rates are introduced by
assuming the same damping rates \cite{mustecap} for the atomic and
photonic modes. The decay rates are obtained from the effective
decay of the experimentally measured contrast for the atomic
density distribution pattern \cite{ketterle}. In addition, we also
explored an interesting scheme where the coupled equations
(\ref{eq:AppendixFirst}-\ref{eq:AppendixLast}) were solved, in the
presence of phenomenological damping, for an initial two-mode
squeezed vacuum (for the end-fire modes) with a squeezing
parameter $\xi=r\exp{(i\theta_r)}$.
\section{Results and Discussion} \label{sec:results}

In sec. \ref{sec:swapmechanism} we discussed the origins of the
entanglement swap in sequential SR. In this section,
in order to provide for a more
detailed and quantitative understanding, we present
results obtained from numerical calculations.
We will discuss the time evolution of the entanglement parameter
$\lambda(t)$, between the two end-fire modes, within the parameter
regime of the experiment \cite{ketterle}. At first, we will
disregard decoherence and examine the nature of
fully coherent sequential dynamics. We will
show that $\lambda$ attains negative values, confirming the
presence of entanglement due to the swap mechanism as we
intuitively discussed in the previous section.
We then introduce effective damping rates
specific to the experimental situation.
Finally, we will examine the effect of
initializing the quantum dynamics of our model system from
in a two-mode (end-fire modes) squeezed vacuum, in the presence of
decoherence and dissipations.
We will end with investigations of the dependence of correlations
on the number of condensate atoms.

\subsection{Dynamics of Entanglement}

\begin{figure}[th]
\includegraphics[width=3.5in]{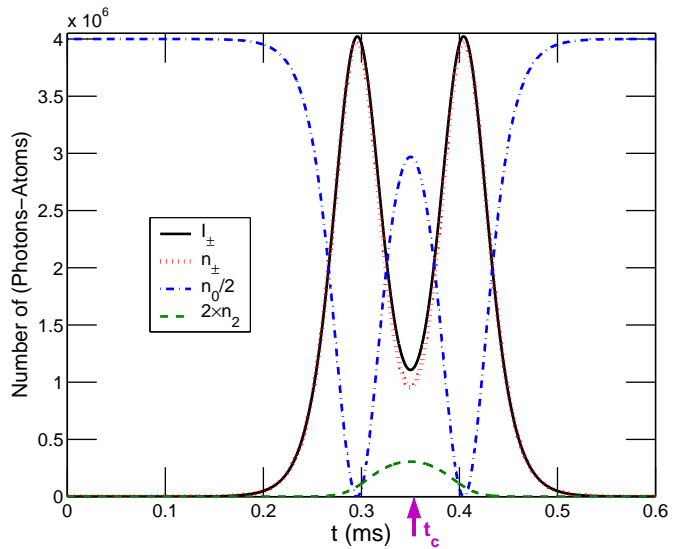}
\caption{(Color online)  The temporal evolutions for atomic
side--mode populations and optical field intensities. $I_\pm$,
$n_\pm$, $n_0$, and $n_2$ denote occupancy numbers of bosonic
modes $|a_\pm\rangle$, $|c_\pm\rangle$, $|c_0\rangle$, and
$|c_2\rangle$, respectively. $n_\pm(t)$ and $I_\pm(t)$ overlap
except for a short time interval near $t=t_c=0.35$ms. Notice that
$n_0$ and $n_2$ are scaled for visual clarity.}
\label{fig2}%
\end{figure}

In Fig. \ref{fig2}, we plot the temporal evolution of optical
field intensities and atomic side--mode populations. The plot is
found to be totally symmetric with respect to $t=t_c=0.35$ ms. The
peak in the intensity after $t_c$ is the analog of the Chiao
ringing \cite{chiao}. In the experiments such a complete ringing
cannot be observed due to the presence of decoherence, which will
be taken into account in our subsequent calculations.

\begin{figure}[th]
\includegraphics[width=3.5in]{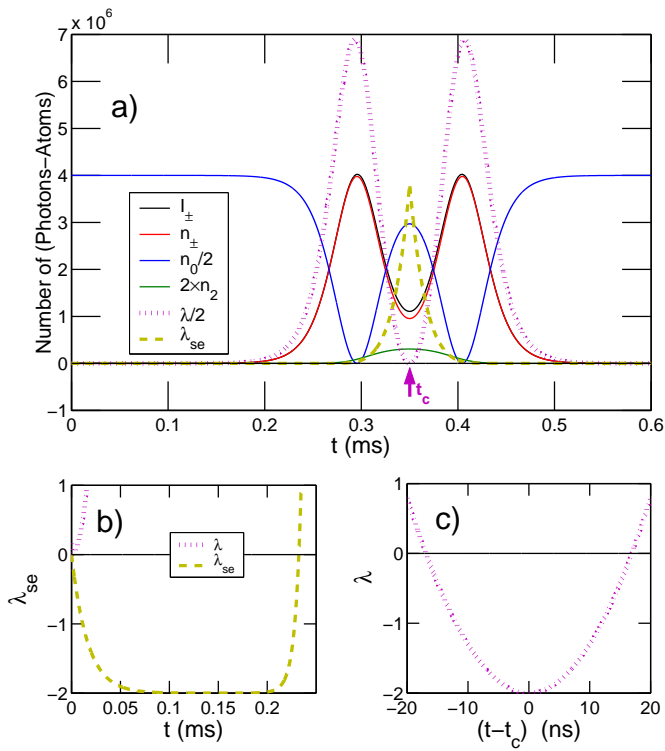}
\caption{(Color online) (a) The temporal evolutions of atom-photon
($|a_\pm\rangle \leftrightarrow |c_\mp\rangle$) and photon-photon
($|a_+\rangle \leftrightarrow |a_-\rangle$) mode correlations as
evidenced by the entanglement parameters $\lambda_{\text{se}}$ and
$\lambda\equiv\lambda_{\text{ee}}$, respectively. Accompanying
population dynamics (Fig. \ref{fig2}) is also plotted in the
background. (b) an expanded view of the early time dynamics for
$\lambda_{\text{se}}$ and $\lambda$; (c) an expanded view of
$\lambda$ around $t_c=0.35$ ms. }
\label{fig3}%
\end{figure}

In Fig. \ref{fig3}, we plot the temporal evolution of entanglement
parameters $\lambda_{\text{ee}}$ (photon-photon) and
$\lambda_{\text{se}}$ (atom-photon) over the  population dynamics,
depicted in Fig. \ref{fig2}. The lower panel of Fig. \ref{fig3}
demonstrates the swap dynamics. The initial atom-photon
entanglement ($\lambda_{\text{se}}$) is seen to evolve
continuously into entanglement between the two end-fire modes
($\lambda$). Both the parameters $\lambda_{\text{se}}$ and
$\lambda$ is found to be able to reach down to the lowest possible
value, $\lambda_{\text{low}}=-2$, set by the Heisenberg
uncertainty principle (as in sec.
\ref{sec:entanglement-criteria}). The complete numerical results
match well with the analytic solutions, discussed previously in
Sec. \ref{sec:swapmechanism}, for the model Hamiltonians
(\ref{eq:H_early}) at early times and (\ref{eq:H_later}) at later
times.

In the time interval of $t=0$-$0.30$ ms, we find the system is
dominated by the $1^{\text{st}}$ sequence of SR. The atomic
condensate, initially in the zero-momentum state $|c_0\rangle$, is
pumped into the $1^{\text{st}}$ order side--modes $|c_\pm\rangle$.
This is the reason for the overlap of $n_\pm(t)$ with $I_\pm(t)$
during this interval. Due to the interaction between the
side--modes and the end-fire modes, scattered into opposite
directions, $\lambda_{\text{se}}$ becomes negative in this region.

When the $|c_\pm\rangle$ side--modes become maximally occupied at
about $t=0.30\,$ms, the $1^{\text{st}}$ sequence of SR is
completed. At this time, these side--modes are sufficiently
populated to give rise to the $2^{\text{nd}}$ sequence of SR. In
the interval $t=0.30$-$0.35\,$ms atoms in the side--modes
$|c_\pm\rangle$ are pumped into the $2^{\text{nd}}$ order
side--mode $|c_2\rangle$. The majority of the populations,
however, oscillates back to the $|c_0\rangle$ mode because of the
more dominant Rabi oscillation between the $|c_0\rangle$ and
$|c_\pm\rangle$ modes. Two other reasons also contribute to the
repopulation of the condensate mode: first, the neglect of the
propagational induced departure of the end-fire mode photons from
the atomic medium; and second, the neglect of the other two
$2^{\text{nd}}$ order side--modes $|c_{2\mathbf{k_0}\pm
2\mathbf{k_e}}\rangle$ for atoms to get into. Two end-fire modes
get indirectly coupled by the entanglement swapping and between
$t=0.30$-$0.35$ ms $\lambda(t)$ gradually becomes negative.

Entanglement of the end-fire modes arises at $t=t_c=0.35$ ms, when
the $|c_2\rangle$ mode is maximally occupied as shown in Fig.
\ref{fig3}. The minimum value of $\lambda$, which occurs at
$t=t_c$, is found to coincide with the maximum value of $n_2(t)$.

When $t>t_c$, however, due to our limited mode approximation of
not including even higher side--modes, we cannot study any effects
which could potential give rise to higher order correlations, such
as the onset of the $3^{\text{rd}}$ sequence of SR. The
oscillatory Chiao type ringing revivals in the present result
after $t>t_c$ mainly arise from the exclusion of decoherence,
dephasing, dissipations, and the higher order side--modes in the
model system. In the present work, we limited ourselves to a
particular side--mode pattern as actually observed in available
experiments \cite{ketterle}. Despite its simplicity, we find our
model can reasonably explain effects of decoherence and dephasing
on the entanglement dynamics, which is further illustrated in the
next subsection.
\subsection{Vacuum squeezing and Decoherence}
\begin{figure}[thb]
\includegraphics[width=3.5in]{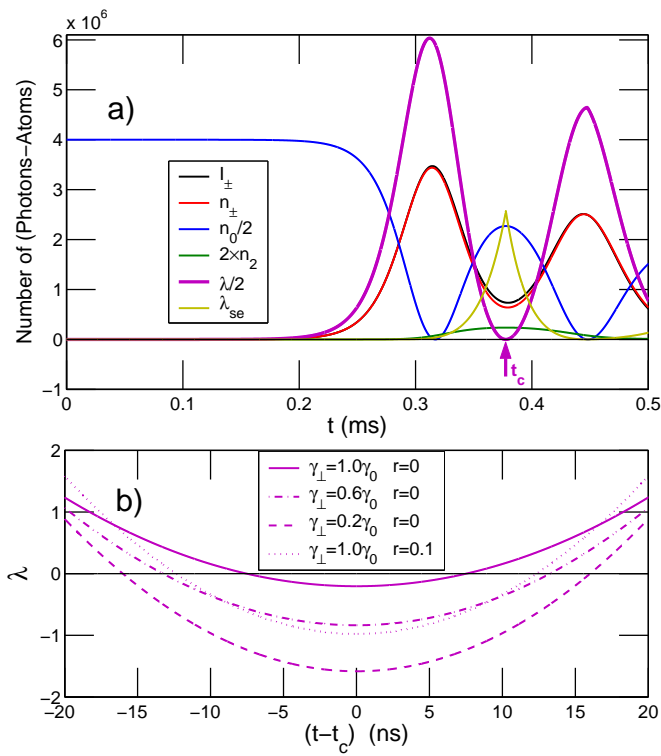}
\caption{(Color online) (a) The temporal evolution of atomic and
field mode populations, and of entanglement parameters. A
decoherence rate of $\gamma_0/2\pi=1.3\times10^4$Hz is introduced
without any initial squeezing. (b) An expanded view of the
dependence of $\lambda$ on decoherence rate $\gamma$ and squeezing
parameter $r$ around $t_c=0.38$ ms.} \label{fig4}
\end{figure}

The introduction of experimentally reported decoherence rate of
$\gamma/2\pi=1.3\times10^4$ Hz phenologically into the dynamical
equations for the coupled system is found to not change the nature
of the entanglement and swap dynamics significantly, which is
supported by the numerical results shown in Fig. \ref{fig4}. We
find that  $\lambda$ can still become negative in certain time
window, although it now stops short of reaching the theoretical
lower bound of $-2$.

In the lower panel in Fig. \ref{fig4} the temporal window for the
negative values of $\lambda$, or the presence of entanglement is
found to become narrower and the minimum value of $\lambda$,
$\lambda_{\text{min}}$, is now somewhat larger for stronger
decoherence, as may be expected. According to sec.
\ref{sec:entanglement-criteria}, a less negative value of
$\lambda$ does not necessarily imply less entanglement, because
$\lambda$ is simply an entanglement witness parameter but not an
entanglement measure. On the other hand, it is still beneficial to
aim for lower values of $\lambda$, because the numerical results
we obtain associate lower values with longer entanglement
durations, and furthermore more tolerant to decoherence, which
means photon-photon entanglement can withstand the higher
decoherence rates.

For this aim, we choose to consider end-fire modes which are
initially in two-mode squeezed vacuum states. The lower panel in
Fig. \ref{fig4} shows that an initially two-mode squeezed vacuum,
for the end-fire modes, can indeed compensate to a certain degree
for decoherence. This shows that, initially induced two-mode
squeezing (or entanglement) in between the end-fire modes enhances
their subsequent entanglement after the entanglement swap.

This observation can be interpreted as follows based on the
numerical results. Any initial correlation between the end-fire
modes is lost in the early dynamical stage where the end-fire
modes are entangled with the first side--modes. The presence of
initial correlation, however, causes the resultant atom-photon
entanglement to be more resistant to decoherence. As a result,
photon-photon correlations established by swapping from the
atom-photon correlations in the subsequent dynamical stage also
become more resistant to decoherence.

\begin{figure}[th]
\includegraphics[width=3.5in]{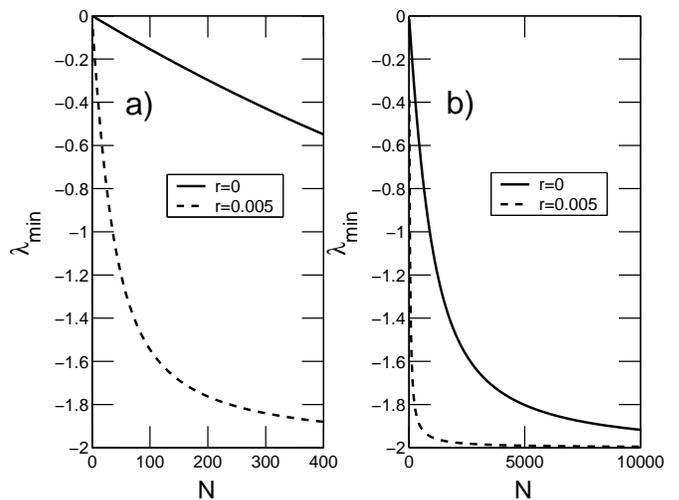}
\caption{The dependence of $\lambda_{\rm min}$ on $N$ in different
scales. Solid lines are for an initial coherent vacuum ($r=0$) and
dashed lines are for a squeezed vacuum ($r=0.005$ and
$\theta=\pi)$.}
\label{fig5}%
\end{figure}

Finally, we examine the influence on $\lambda$ from the number of
atoms in a condensate. We find that, as illustrated in Fig.
\ref{fig5}, $\lambda_{\text{min}}$ becomes more negative for
larger condensates. In the small condensate limit,
$\lambda_{\text{min}}$ is found to decrease linearly with $N$ when
the Fock vacuum is considered as initial conditions for other
modes. The lower limit of $-2$ is never attained. When a small
amount of initial squeezing is introduced, however, $\lambda$ can
be brought down to theoretical minimum of $-2$. It approaches $-2$
in the large condensate limit with or without any help from
initial squeezing in the two end-fire modes.

\begin{figure}[th]
\includegraphics[width=3.5in]{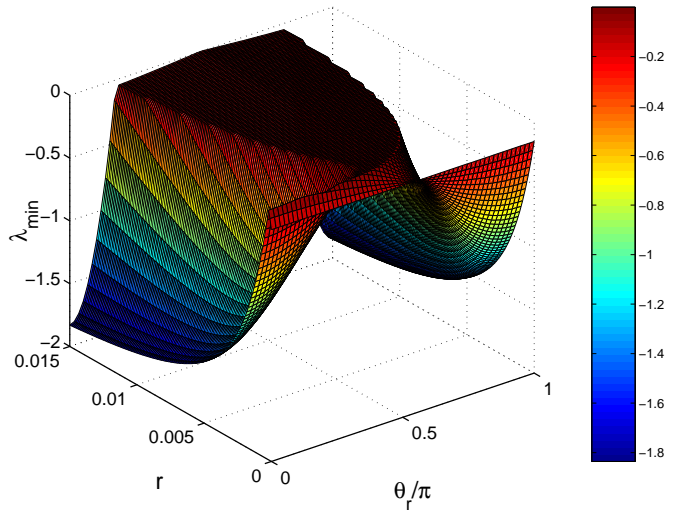}
\caption{(Color online) The dependence of $\lambda_{\rm min}$ on
$r$ and $\theta_r$ for $N=100$. $\lambda_{\rm min}$ shows a mirror
symmetry for $\theta_r>\pi$.}
\label{fig6}%
\end{figure}

In addition to the amplitude of squeezing parameter, its phase
could also influence $\lambda_{\rm min}$. In Fig. \ref{fig6}, we
plot the minimum value of the entanglement parameter as a function
of the phase and amplitude of the squeezing parameter
$\xi=re^{i\theta_r}$. We performed this study for a small
condensate with $N=100$ atoms and ignored the phenological
decoherence. We find that the most efficient enhancement occurs
along the line $\theta_r=\pi$. For larger condensates we find that
the center of Fig. \ref{fig6}, where $\lambda>0$, spreads out to
the $\theta_r=0$ and $\theta_r=\pi$ edges, as $N$ is increased.
Entanglement is enhanced mainly along $\theta_r=0$ and
$\theta_r=\pi$ lines.

%
\section{Conclusions} \label{sec:conclusions}

We investigate photon-photon entanglement between the
counter-propagating end-fire modes, of a sequentially superradiant
atomic Bose-Einstein condensate. We calculate the temporal
evolution of the continuous variable entanglement witness
parameter for suitable realistic experimental parameters in the
cw-pump laser regime \cite{ketterle}, and find that EPR like
correlations can be generated between the oppositely directed
end-fire modes, despite the fact that they do not directly
interact.

The generation of entanglement is shown to be due to a built-in
entanglement swap mechanism we uncover in the sequential SR
system. It is shown that end-fire mode photons become entangled
immediately after the second sequence of the superradiance. In the
second sequence, one of the end-fire modes interacts with the
side--mode, with which the other end-fire mode has already
interacted before in the first sequence. This mechanism allows for
swapping the entanglement established between the end-fire modes
and the side--modes in the first sequence, to the entanglement of
the end-fire modes per se.

Increasing the number of atoms in the condensate, or initializing
superradiance with a two-mode squeezed vacuum (for the end-fire
modes), are found to be beneficial to the efficient construction of
entanglement between end-fire modes, via the increasing of
entanglement durations and making the entanglement more tolerant to
decoherence.

The initial phase difference of the incoming pump laser and the
condensate, the phase and the amplitude of the squeezing parameter
for the end-fire mode vacuum, the number of atoms in a condensate
and its geometric shape, all play certain roles in order to
achieve the optimum ERP-like correlations in between the end-fire
modes and these parameters are discussed in detail in the present
manuscript for the cases of both small and large condensates.
\begin{acknowledgements}
M.\"O.O. is supported by a T\"UBA/GEB\.{I}P grant and
T\"UB\.ITAK-KAR\.IYER Grant No. 104T165. L.Y. acknowledges the
support from US NSF and ARO.
\end{acknowledgements}
\appendix
\section{}
We calculate temporal evolution of entanglement parameter
$\lambda(t)$, given in (\ref{eq:lambda-endfires}),
starting from the
Heisenberg operator equations, obtained from (\ref{eq:hamiltonian}).
We
evaluate the expectations for both single operators and two operator
products. We arrive at a closed set from the expectations via
performing decorrelation approximation, in parallel with the
development and understanding of the swap
mechanism (Sec. \ref{sec:swapmechanism}).

The resulting closed set of equations for expectation values are
given through Eqs. (\ref{eq:AppendixFirst}-\ref{eq:AppendixLast}),
where time is scaled by frequency $\alpha=g^2/2\Delta$ with
$g\simeq2\times 10^3$ Hz, while operators are not scaled. $\alpha$
is related to $\gamma$ of Ref. \cite{mustecap}, as
$\gamma=\sqrt{M}\alpha=10.7$ Hz. Phenological decay rates can be
introduced in Eqs. (\ref{eq:AppendixFirst}-\ref{eq:AppendixLast})
by scaling $\gamma_\perp=1.3\times 10^4$Hz with $\alpha$. However,
since the decay rates are introduced, in \cite{mustecap}, for
three-operator products, we use $\gamma_\perp/3$ for single
operators and $2\gamma_\perp/3$ for two operator products. We have
also checked the parallelism of our density dynamics with
\cite{mustecap}, which are in good agreement with the experiment
\cite{ketterle}.

\begin{eqnarray}
\frac{d\langle a_\pm\rangle}{dt} &=& i \langle a_0 \rangle \Big(
\langle c_\mp^\dagger \rangle \langle c_0\rangle + \langle
c_2^\dagger \rangle \langle c_\pm\rangle \Big),
\label{eq:AppendixFirst}
\\
\frac{d\langle a_0 \rangle}{dt}&=&i \Big( \langle a_-c_+\rangle
\langle c_0^\dagger \rangle + \langle a_-c_-^\dagger \rangle \langle
c_2 \rangle
\nonumber \\
&+& \langle a_+c_-\rangle \langle c_0^\dagger \rangle + \langle
a_+c_+^\dagger\rangle \langle c_2 \rangle \Big),
\\
\frac{d\langle c_\pm \rangle}{dt} &=& i \Big( \langle a_0 \rangle
\langle a_\mp^\dagger\rangle \langle c_0\rangle + \langle a_0
\rangle^* \langle a_\pm\rangle \langle c_2\rangle \Big),
\\
\frac{\langle c_0\rangle}{dt} &=&i \langle a_0 \rangle^* \Big(
\langle a_-c_+ \rangle + \langle a_+c_-\rangle \Big),
\\
\frac{\langle c_2\rangle}{dt} &=& i \langle a_0 \rangle \Big(
\langle a_-^\dagger c_- \rangle + \langle a_+^\dagger c_+\rangle
\Big),
\\
\frac{d\langle a_+a_-\rangle}{dt} &=& i \langle a_0 \rangle \Big(
\langle a_-c_-^\dagger\rangle \langle c_0\rangle + \langle
a_-c_+\rangle \langle c_2^\dagger\rangle \nonumber \\ &+& \langle
a_+c_+^\dagger \rangle \langle c_0\rangle + \langle a_+c_-\rangle
\langle c_2^\dagger\rangle \Big),
\\
\frac{d\langle a_+ a_-^\dagger \rangle}{dt} &=& i\Big( \langle a_0
\rangle \langle a_-^\dagger c_-^\dagger \rangle \langle c_0 \rangle
+\langle a_0 \rangle \langle a_-^\dagger c_+ \rangle \langle
c_2^\dagger
\rangle \nonumber \\
&-&\langle a_0 \rangle^* \langle a_+c_+ \rangle \langle c_0^\dagger
\rangle +\langle a_0 \rangle^* \langle a_+c_-^\dagger \rangle
\langle c_2 \rangle \Big),
\\
\frac{d\langle a_\pm^2 \rangle}{dt} &=& 2i \langle a_0 \rangle \Big(
\langle a_\pm c_\mp^\dagger  \rangle \langle c_0 \rangle +\langle
a_\pm c_\pm \rangle \langle c_2^\dagger \rangle \Big),
\\
\frac{d\langle c_+ c_-\rangle}{dt} &=& i \Big( \langle a_0 \rangle
\langle a_-^\dagger c_-\rangle \langle c_0 \rangle +\langle a_0
\rangle^* \langle a_+c_- \rangle \langle c_2\rangle \nonumber \\&+&
\langle a_0 \rangle \langle a_+^\dagger c_+ \rangle \langle
c_0\rangle +\langle a_0 \rangle^* \langle a_-c_+ \rangle \langle c_2
\rangle \Big),
\\
\frac{d\langle c_+ c_-^\dagger \rangle}{dt} &=& i \Big( \langle a_0
\rangle \langle a_-^\dagger c_-^\dagger \rangle \langle c_0 \rangle
+\langle a_0 \rangle^* \langle a_+c_-^\dagger \rangle \langle
c_2\rangle \nonumber \\ &-&\langle a_0 \rangle^* \langle a_+c_+
\rangle \langle c_0^\dagger \rangle -\langle a_0 \rangle \langle
a_-^\dagger c_+ \rangle \langle c_2^\dagger \rangle \Big),
\\
\frac{\langle c_\pm^2 \rangle}{dt} &=& 2i \Big( \langle a_0 \rangle
\langle a_\mp^\dagger c_\pm \rangle \langle c_0 \rangle
\nonumber\\&+&\langle a_0 \rangle^* \langle a_\pm c_\pm \rangle
\langle c_2\rangle \Big),
\\
\frac{d\langle a_\pm c_\mp \rangle}{dt} &=& i \Big( \langle a_0
\rangle \langle c_\mp^\dagger c_\mp \rangle \langle c_0\rangle +
\langle a_0 \rangle \langle c_\pm c_\mp \rangle \langle
c_2^\dagger\rangle \nonumber \\ &+& \langle a_0 \rangle \langle
a_\pm a_\pm^\dagger \rangle \langle c_0\rangle + \langle a_0
\rangle^* \langle a_\pm a_\mp \rangle \langle c_2\rangle \Big),
\\
\frac{d\langle a_\pm c_\pm^\dagger \rangle}{dt} &=& i \Big( \langle
a_0 \rangle \langle c_\pm^\dagger c_\mp^\dagger \rangle \langle c_0
\rangle + \langle a_0 \rangle \langle c_\pm c_\pm^\dagger \rangle
\langle c_2^\dagger \rangle \nonumber \\ &-& \langle a_0 \rangle^*
\langle a_\pm a_\mp \rangle \langle c_0^\dagger\rangle - \langle a_0
\rangle^* \langle a_\pm a_\pm^\dagger \rangle \langle c_2^\dagger
\rangle \Big),
\\
\frac{\langle a_\pm c_\pm \rangle}{dt} &=& i \Big( \langle a_0
\rangle \langle c_\pm c_\mp^\dagger \rangle \langle c_0\rangle
+\langle a_0 \rangle \langle c_\pm^2 \rangle \langle
c_2^\dagger\rangle \nonumber \\ &+&\langle a_0 \rangle \langle a_\pm
a_\mp^\dagger \rangle \langle c_0 \rangle +\langle a_0 \rangle^*
\langle a_\pm^2 \rangle \langle c_2 \rangle \Big),
\\
\frac{\langle c_\pm^\dagger c_\pm\rangle}{dt} &=&
i \Big( -\langle a_0 \rangle^* \langle a_\mp c_\pm\rangle \langle
c_0^\dagger\rangle -\langle a_0 \rangle \langle a_\pm^\dagger c_\pm
\rangle \langle c_2^\dagger\rangle \nonumber \\ &+&\langle a_0
\rangle \langle a_\mp^\dagger c_\pm^\dagger \rangle \langle
c_0\rangle + \langle a_0 \rangle^* \langle a_\pm c_\pm^\dagger
\rangle \langle c_2 \rangle \Big),
\\
\frac{\langle c_0^\dagger c_0 \rangle}{dt} &=& i \Big( -\langle a_0
\rangle \langle a_-^\dagger c_+^\dagger \rangle \langle c_0 \rangle
-\langle a_0 \rangle \langle a_+^\dagger c_-^\dagger \rangle \langle
c_0 \rangle \nonumber \\ &+&\langle a_0 \rangle^* \langle a_-c_+
\rangle \langle c_0^\dagger \rangle +\langle a_0 \rangle^* \langle
a_+c_- \rangle \langle c_0^\dagger \rangle \Big),
\\
\frac{\langle c_2^\dagger c_2 \rangle}{dt} &=& i \Big( -\langle a_0
\rangle^* \langle a_-c_-^\dagger \rangle \langle c_2 \rangle
-\langle a_0 \rangle^* \langle a_+c_+^\dagger \rangle \langle c_2
\rangle \nonumber \\ &+&\langle a_0 \rangle \langle a_-^\dagger c_-
\rangle \langle c_2^\dagger \rangle +\langle a_0 \rangle \langle
a_+^\dagger c_+ \rangle \langle c_2^\dagger \rangle \Big),
\\
\frac{\langle a_\pm^\dagger a_\pm\rangle}{dt} &=& i \Big( -\langle
a_0 \rangle^* \langle a_\pm c_\mp\rangle \langle c_0^\dagger\rangle
-\langle a_0 \rangle^* \langle a_\pm c_\pm^\dagger \rangle \langle
c_2\rangle \nonumber \\ &+& \langle a_0 \rangle \langle
a_\pm^\dagger c_\mp^\dagger \rangle \langle c_0\rangle + \langle a_0
\rangle \langle a_\pm^\dagger c_\pm \rangle \langle c_2^\dagger
\rangle \Big). \label{eq:AppendixLast}
\end{eqnarray}


\end{document}